\documentclass[amsmath,twocolumn,showpacs,amssymb,aps,nofootinbib]{revtex4-2}
\usepackage{graphicx}
\usepackage{subcaption}
\usepackage{pgfplots}
\usepackage{dcolumn}
\usepackage{bm}
\usepackage{slashed}
\usepackage{amsmath}
\usepackage{amssymb}
\usepackage{tikz}
\usepackage{bbold}
\usepackage{mathtools}
\usepackage{romannum}
\usepackage{natbib}
\usepackage{braket}
\usepackage{appendix}
\begin{document}
\pagenumbering{arabic}

\title{FOTOC complexity in an extended Lipkin-Meshkov-Glick model \\}
\author{Nitesh Jaiswal}
\email{nitesh@iitk.ac.in}
\author{Mamta Gautam}
\email{mamtag@iitk.ac.in}
\author{Ankit Gill}
\email{ankitgill20@iitk.ac.in}
\author{Tapobrata Sarkar}
\email{tapo@iitk.ac.in}
\affiliation{
Department of Physics, Indian Institute of Technology Kanpur-208016, India}
\begin{abstract}
We study fidelity out-of-time-order correlators (FOTOCs) in an extended
Lipkin-Meshkov-Glick model and demonstrate that these exhibit distinctive behaviour
at quantum phase transitions in both the ground and the
excited states. We show that the dynamics of the FOTOC have different behaviour
in the symmetric and broken-symmetry phases, and as one approaches phase transition.
If we rescale the FOTOC operator with time, then for small times, we establish
that it is identical to the Loschmidt echo. We also compute the Nielsen complexity 
of the FOTOC operator in both phases, and apply this operator on the ground and excited
states to obtain the quasi-scrambled state of the model. The FOTOC
operator introduces a small perturbation on the original ground and
excited states. For this perturbed state, we compute the quantum information metric to
first order in perturbation, in the thermodynamic limit. We find that the associated Ricci scalar 
diverges at the phase transition on the broken-symmetry phase side, in contrast to the
zeroth order result. Finally, we comment upon the Fubini-Study complexity in this model.
\end{abstract}
\maketitle
\section{\label{introduction}Introduction}
Out-of-time-order correlators (OTOCs) were originally introduced in the context of vertex corrections in superconductors \cite{Larkin} and
currently find widespread usage as a measure of quantum chaos, with its growth rate being related to the Lyapunov exponent 
\cite{Chavez,Akutagawa,Hashimoto,Rozenbaum}. The OTOC is also the key quantity in studies of quantum information geometry
(QIG), in particular in quantum information 
scrambling \cite{Swingle1, Hauke, Harrow} and quantum phase transitions (QPTs) \cite{Shen, Wang, Kidd, Heyl, Shukla, Huh, Rivera}. 
Depending on how the OTOC is computed, it may be referred to as the microcanonical OTOC, the fidelity OTOC (FOTOC), 
the thermal OTOC, etc. In quantum systems, the OTOC may serve as a diagnostic tool to probe ground-state QPTs (GSQPTs) 
and excited-state QPTs (ESQPTs). Another reason why the FOTOC is important is that the OTOC features 
exponential growth at unstable points or in chaotic regions of the quantum system. However, exact numerical treatment 
is only possible for a small system size $N$, as many-body observables quickly saturate with time, $t\sim\log N$.
This limitation can be overcome if one uses the FOTOC \cite{Cameo,Swan}. One purpose of this paper is to examine 
the FOTOC as a probe for QPTs in QIG, as it can be directly implemented 
and measured in experiments with trapped ions \cite{Swan,Naini}. The detailed experimental studies and demonstrations to measure OTOCs are performed in \cite{Joshi, Green, Landsman, Garttner}.

On the other hand, scrambling is the process by which information stored in local degrees of freedom is dispersed 
across a large number of degrees of freedom. The Heisenberg picture, in which quantum operators evolve and quantum states are 
stationary, contains a precise formulation of quantum scrambling. The structure of the time-dependent Heisenberg 
operator $\hat{W}(t)$ for any operator  $\hat{W}$ resembles a classical butterfly effect. Even if $\hat{W}$ is a 
local operator, such as a spin operator at a given site, $\hat{W}(t)$ will eventually spread across many sites, making it 
inaccessible to local measures \cite{Swingle, Mi, Swingle2, Swan}. There are two distinct types of scramblings known in 
the current literature \cite{Zhuang, Yan}, which are termed as {\it genuine scrambling}, where a localised initial operator in 
phase space spreads significantly, and {\it quasi-scrambling}, where a localised initial operator stays localised, but can 
nonetheless move around in phase space. Even if quasi-scramblers are not fully spread operators, useful insights into 
genuine scrambling can be obtained by their study. A second purpose of this paper is to study the QIG of quasi-scrambled
states. 

Indeed, the primary reason for using FOTOC is that it allows us to visualize the dynamics of
scrambling using a semi-classical picture. The FOTOC can be mapped to a two-point
correlator, allowing us to compute it in a parameter regime inaccessible to exact
numerical diagonalisation using phase-space methods. Secondly, the FOTOC has the
potential to shed light on QPTs. FOTOC's time evolution have distinct behaviours in
different phases, whereas its behaviour on QPTs is very different in comparison
to the these phases. Unlike microcanonical OTOCs, where we need to choose order parameters
as our operators to identify the existence of QPTs, for FOTOCs we just need to select
the Hermitian operator.

With these motivations to study the above mentioned information theoretic quantities, 
in this paper we consider an extension of the Lipkin-Meshkov-Glick 
model which we call the eLMG model, introduced in \cite{GGCHV}, which is attractive, as it allows very efficient analytical calculations 
and numerical treatments much like the original LMG model. The LMG model which is exactly solvable, was introduced in the context of nuclear 
physics \cite{LMG}, and is an ideal arena to explore effects of long range interactions (beyond the nearest neighbour
ones which are well studied for example in the transverse field XY model) which result in quantum phase 
transitions in the thermodynamic limit \cite{Botet1},\cite{Botet2}.  
In the context of the eLMG model as well, analytical calculations are simplified, since the model reduces to a
simple harmonic oscillator after a Bogoliubov transformation. The numerical
calculations too are straightforward here, since the size of the Hamiltonian
matrix to be diagonalised grows linearly with the number of spins $N$, if we consider
the maximum spin sector $j=N/2$, which contains low energy states. The eLMG model
is considered here rather than the standard LMG model, since it was shown in \cite{tapo5} that 
information geometry is ill-defined in the latter. In contrast, QIG is well defined in the eLMG model.

In this work, we find the Nielsen complexity (NC) \cite{Qu, Arpan1, Arpan, Choudhury, Liu, Haque, tapo1, 
tapo2, tapo3, tapo4, Guo} of the time-dependent FOTOC operator $\hat{W}(t)$ in an eLMG model, and will show that the derivative 
of the NC diverges, if we approach the QPT line from the broken-symmetry phase side. We also apply the time-dependent 
FOTOC operator on the ground and excited state to get the quasi-scrambled state. We then compute the 
quantum information metric (QIM) \cite{tapo1, tapo2, tapo3, tapo4, Miyaji} for these quasi-scrambled states and
study its various properties. Natural units with $\hbar = 1$ are used throughout. 

\section{\label{model}The Hamiltonian and its Semiclassical description}

We now briefly review some of the necessary formalism of \cite{GGCHV}. The Hamiltonian of an eLMG model is 
\begin{equation}
H=\Omega \hat{J}_{z}+\Omega_{x}\hat{J}_{x}+\frac{\xi_{y}}{j}\hat{J}^{2}_{y}~,\label{Hamilt}
\end{equation}
where $\hat{J}_{\alpha}=\sum_{i=1}^{N}\sigma_{\alpha}^{i}$ are the collective pseudospin operators, 
and $\sigma_{\alpha}^{i}$ are the Pauli spin operators for a particular two-level system $i$ along $\alpha=x,y,z$ directions. The parameter $\Omega$ gives the energy difference for the two-level system, and from now on we will set $\Omega=1$. 
The parameter space of interest is $\Omega_{x}$, $\xi_{y}$ $\in$ $\mathbb{R}$, where $\Omega_{x}$ is the strength of the linear term included in the LMG model to transform it to the eLMG model and $\xi_{y}$ is the coupling strength.
The collective spin operators follow the commutation relation $[ \hat{J}_{\alpha} ,\hat{J}_{\beta}]= 
i\epsilon_{\alpha\beta\gamma} \hat{J}_{\gamma}$, and the magnitude of total spin operator is conserved, i.e., 
$[H,\hat{J}^{2}]=0$, with $j(j+1)$ being the eigenvalue of $\hat{J}^{2}$. The $j$ in the denominator 
of Eq. (\ref{Hamilt}) ensures finite energy per spin in the thermodynamic limit, and as mentioned in the introduction, here for 
simplicity, we consider the maximum spin sector by fixing $j=N/2$ to get the size of the system. 
The classical eLMG Hamiltonian is obtained 
by taking its expectation value with respect to Bloch coherent states $\ket{z}=(1+|z|^{2})^{-j}e^{z\hat{J}_{+}}\ket{j,-j}$.
Here, $\hat{J}_{+}$ is the raising operator, and $\ket{j,-j}$ is the state with the lowest pseudospin projection, 
and $z$ $\in$ $\mathbb{C}$ is defined in terms of spherical polar coordinates, $z=\tan\left(\frac{\theta}{2}\right)
e^{-i\phi}$ with semiclassical polarisation,
\begin{equation}
\braket{\hat{J}}=j\left(\sin\theta\cos\phi,\sin\theta\sin\phi,\cos\theta\right)~.
\end{equation}
The classical eLMG Hamiltonian has the form
\begin{equation}
h=-\cos\theta+\Omega_{x}\sin\theta\cos\phi+\xi_{y}\sin^{2}\theta\sin^{2}\phi~,
\end{equation}
which can also be written using convenient canonical variables $Q$ and $P$, defined as
\begin{equation}
Q=\sqrt{2(1-\cos\theta)}\cos\phi,\,P=-\sqrt{2(1-\cos\theta)}\sin\phi~.
\end{equation}
With the classical Hamiltonian, we can obtain the stationary points of the eLMG model, which corresponds to vanishing velocities 
\begin{equation}
\dot{Q}=\frac{\partial h}{\partial P}=0,\;\dot{P}=-\frac{\partial h}{\partial Q}=0~,
\end{equation}
in the region $0\leq\theta\leq\pi$ and $0\leq\phi\leq2\pi$. The stationary points under these conditions are
\begin{eqnarray}
(\theta_{1},\phi_{1})&=&\left(\arccos\left(\frac{1}{\sqrt{\Omega_{x}^{2}+1}}\right),0\right),\notag\\
(\theta_{2},\phi_{2})&=&\left(\arccos\left(\frac{1}{\sqrt{\Omega_{x}^{2}+1}}\right),\pi\right)~.
\end{eqnarray}
Another stationary point $(\theta_{3},\phi_{3})$ that was given in \cite{GGCHV} is only valid for 
$\xi_{y}\leq-\sqrt{1+\Omega_{x}^{2}}/2$. In what follows, we consider the domain $\xi_{y}\geq0$, and will 
only illustrate the results for $\xi_{y}\leq0$ whenever necessary. Finally, the stationary point 
\begin{equation}
(\theta_{4},\phi_{4})=\left(\arccos\left(-\frac{1}{2\xi_{y}}\right),\arccos
\left(\frac{\Omega_{x}}{\sqrt{4\xi_{y}^{2}-1}}\right)\!\!\right)~,
\end{equation} 
is valid for $\xi_{y}\geq\sqrt{1+\Omega_{x}^{2}}/2$. 

The energies $e_{n}=h(\theta_{n},\phi_{n})$ associated with the stationary points are 
$e_{1,2}=\pm\sqrt{1+\Omega_{x}^{2}}$, and $e_{3}=e_{4}=(1+\Omega_{x}^{2})/(4\xi_{y})+\xi_{y}$. 
Note that the energies $e_{3}$ and $e_{4}$ are defined only in the regions 
$\xi_{y}\leq-\sqrt{1+\Omega_{x}^{2}}/2$ and $\xi_{y}\geq\sqrt{1+\Omega_{x}^{2}}/2$, respectively,
as follows from our previous discussion.  
The GSQPT line, defined as $\xi_{y}=-\sqrt{1+\Omega_{x}^{2}}/2$, separates the two quantum phases 
of the ground state, described on one side by $(\theta_{2},\phi_{2})$ and on the other by 
$(\theta_{4},\phi_{4})$ coherent states. Similarly, the ESQPT line, denoted by $\xi_{y}=
\sqrt{1+\Omega_{x}^{2}}/2$, acts as a dividing line between the quantum phases of the excited state 
described by coherent states $(\theta_{1,4},\phi_{1,4})$ on the two sides.

\section{\label{fotoc}FOTOC and Quantum Phase Transitions}

We now examine the GSQPT and ESQPT of an eLMG model using FOTOCs. Typical usage identifies the 
OTOCs as measures of the dynamics of quantum information scrambling. Concurrently, OTOCs are defined as 
$F(t)=\braket{\hat{W}^{\dagger}(t)\hat{V}^{\dagger}\hat{W}(t)\hat{V}}$, where $\hat{V}$ and $\hat{W}$ 
are two Hermitian operators, $\hat{W}(t)=e^{i\hat{H}t}\hat{W}e^{-i\hat{H}t}$ is the time-dependent operators 
at time $t$ in the Heisenberg representation, with $\hat{H}$ being the Hamiltonian of the system, and 
$\braket{\cdots}$ represents thermal averaging. This OTOC is called FOTOC \cite{Swan, Cameo} if we only allow 
a small perturbation $\varepsilon\ll 1$ to the operator $\hat{W}=e^{i\varepsilon \hat{G}}$, where $\hat{G}$ 
is a Hermitian operator and sets $\hat{V}=\ket{\psi_{in}}\bra{\psi_{in}}$ to be a projection operator onto an 
initial state $\ket{\psi_{in}}$. Then for the pure state, $F(t)\equiv|\bra{\psi_{in}}\hat{W}(t)\ket{\psi_{in}}|^{2}$ 
is known as FOTOC and up to $\mathcal{O}(\varepsilon^{2})$, it reduces to $F_{G}(t)\approx1-\varepsilon^{2}
\sigma_{G}^{2}(t)$ where
\begin{equation}
\sigma_{G}^{2}(t)=\left(\bra{\psi_{in}}\hat{G}^{2}(t)\ket{\psi_{in}}-\bra{\psi_{in}}\hat{G}(t)\ket{\psi_{in}}^{2}\right)~.
\end{equation}
This last relation directly links the FOTOC and the two-point correlator. This also maps 
it to the variances of the operator $\hat{G}(t)$, i.e., the FOTOC can be viewed as the square of the 
uncertainty in the operator $\hat{G}(t)$ in the initial state up to order $\varepsilon^{2}$. 
In our generic treatment to compute the FOTOC, we choose the operator $\hat{G}=\hat{Q}$, $\hat{P}$, 
which are position and momentum operators respectively of the form 
\begin{equation}
\hat{Q}=\frac{1}{\sqrt{2}}(\hat{a}+\hat{a}^{\dagger}),\;\hat{P}=\frac{i}{\sqrt{2}}
(\hat{a}^{\dagger}-\hat{a})~,\label{QPtransform}
\end{equation}
and $\hat{V}=\ket{z}\bra{z}$, which is a projection operator on the spin coherent state. By selecting the 
position and momentum operators, the FOTOC measures the spread of the wavepacket's size, depicting 
quantum evolution of the phase space dynamics. Since the wavepacket can spread in either way in phase space, 
we must examine the growth of $F_{Q}(t)+F_{P}(t)$. Note that in order to operate $\hat{a}$ and $\hat{a}^{\dagger}$ on spin 
coherent states, we first write them in terms of $\hat{J}_{\pm}=\hat{J}_{x}\pm i\hat{J}_{y}$: $\hat{a}=J_{-}/\sqrt{2j}$ 
and $\hat{a}^{\dagger}=J_{+}/\sqrt{2j}$.
\begin{figure}[h!]
\centering
\includegraphics[width=0.47\textwidth]{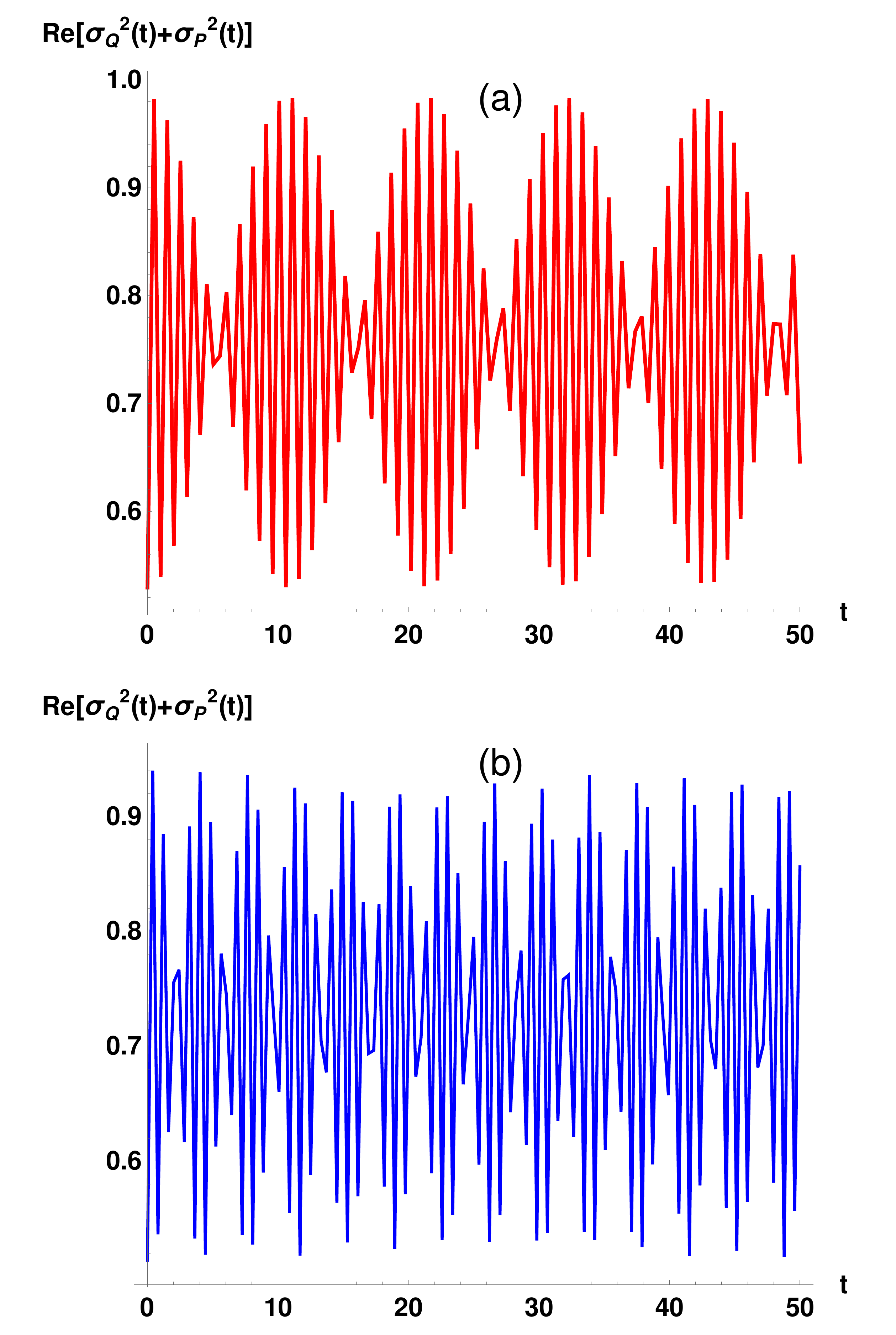}
\caption{Comparison between the behavior of the FOTOC's real part versus time in the 
symmetric phase $(\theta_{1},\phi_{1})$, $\xi_{y}=1$ (a), and broken-symmetry phase 
$(\theta_{4},\phi_{4})$, $\xi_{y}=3$ (b), with $j=400$ and $\Omega_{x}=4$ of the excited state. 
The difference in the oscillatory nature of the two phases is apparent from the top and bottom panel.}
\label{otocsb}
\end{figure}

\begin{figure}[h!]
\centering
\includegraphics[width=0.47\textwidth]{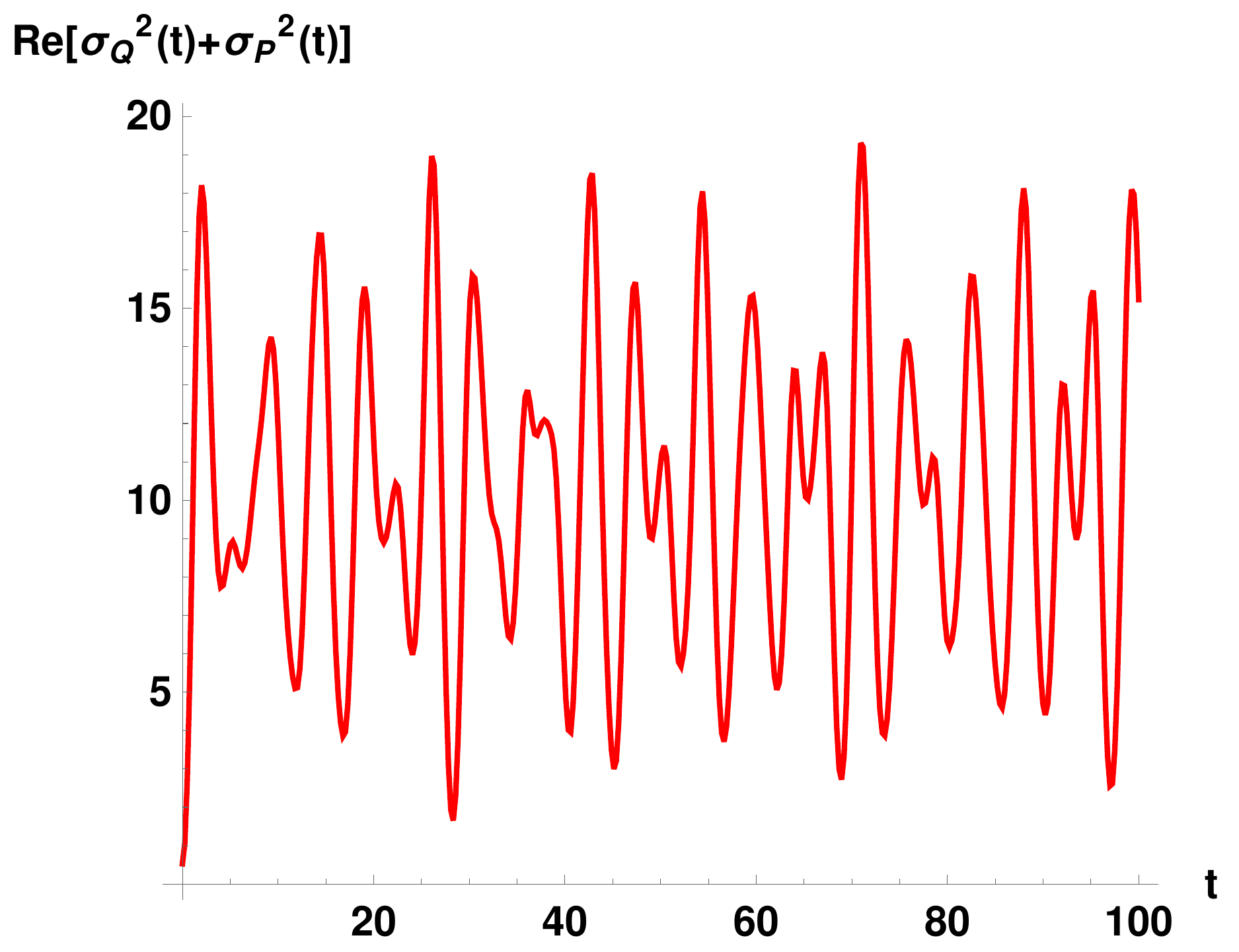}
\caption{The temporal evolution of the FOTOC at ESQPT is described by $\xi_{y}=\sqrt{1+\Omega_{x}^{2}}/2$ 
with $j= 400$ and $\Omega_{x}=4$. The oscillations are irregular, with much higher amplitudes as 
compared to symmetric and broken-symmetry phases.}
\label{otocqpt}
\end{figure}

In Fig. (\ref{otocsb}), we have presented the numerical results of the FOTOC evaluated for an eLMG model. 
The time evolution of the real part of the FOTOC computed for the symmetric phase $(\theta_{1},\phi_{1})$, $\xi_{y}=1$ 
in Fig. \ref{otocsb}(a), broken-symmetry phase $(\theta_{4},\phi_{4})$, $\xi_{y}=3$,  in Fig. \ref{otocsb}(b), 
with $j=400$, and $\Omega_{x}=4$ has been shown here. As shown in Fig. \ref{otocsb}(a), the evolution of the 
FOTOC in the symmetric phase consists of a sequence of typical wave packets. The envelope of a wave packet 
is a measure of the variance of the sum of the position and momentum operators, which extends over a finite time. 
The pattern of the wave packet remains the same, even for a long time, as we have checked for time values up to 5000. 
In contrast with the symmetric phase, the time evolution of the FOTOC in the broken-symmetry phase consists of a 
series of continuous oscillations with a mixture of small and large amplitudes. The oscillations persist here 
for large times as well. In Fig. (\ref{otocqpt}), we plot the evolution of the FOTOC on the QPT line 
$\xi_{y}=\sqrt{1+\Omega_{x}^{2}}/2$, where, after a rapid increase, the FOTOC begins oscillating with an irregular 
pattern of high amplitude. These oscillations do not die out with time even at QPTs, contrary to the microcanonical 
OTOC's in \cite{Wang, Heyl}. In conclusion, the ESQPT is characterised by the FOTOC's abrupt increase in amplitude 
and irregular oscillations, while the oscillatory patterns in the two phases are very distinct.

It is worth noting that a similar situation arises for the GSQPT. We have repeated the analysis above 
for the ground state, and find that GSQPT occurs at $\xi_{y}=-\sqrt{1+\Omega_{x}^{2}}/2$, as evidenced 
by an increase in amplitude with irregular oscillations. One phase of the ground state is described 
by $(\theta_{2},\phi_{2})$ for $\xi_{y}>-\sqrt{1+\Omega_{x}^{2}}/2$ and this has the FOTOC's temporal 
evolution similar to what we found for the symmetric phase of the ESQPT. The other phase is described 
by $(\theta_{4},\phi_{4})$, the analysis of which was done in the excited state case. In addition, the 
FOTOC also provides a link between the classical and quantum Lyapunov exponents \cite{Cameo, Swan}. The eLMG 
model has unstable points $(\theta_{1},\phi_{1})$ for $\xi_{y}>\sqrt{1+\Omega_{x}^{2}}/2$ and 
$(\theta_{2},\phi_{2})$ for $\xi_{y}<-\sqrt{1+\Omega_{x}^{2}}/2$. 
This gives rise to a positive classical Lyapunov exponent given by
\begin{equation}
\lambda_{cl}=\sqrt{\sqrt{1+\Omega_{x}^{2}}(2\xi_{y}-\sqrt{1+\Omega_{x}^{2}})}~,
\end{equation}
where the FOTOC exhibits exponential growth, i.e., $1-\text{Re}[F_{G}(t)]\sim e^{\lambda_{Q}t}$. 
Here, $\lambda_{Q}$ is the quantum Lyapunov exponent and we have checked numerically 
that $\lambda_{Q}\simeq2\lambda_{cl}$.

\subsection{\label{fotoc1}Loschmidt Echo and the FOTOC}

The LE, denoted by $\mathcal{L}$ (see, e.g. \cite{tapo2}), quantifies how similar a state is to one that 
has been time-evolved by a Hamiltonian $\hat{H}_{F}$ and then backward in time by a slightly different 
Hamiltonian $\hat{H}$. We will take the state to be the spin coherent state, the Hamiltonian $\hat{H}=H$ 
and $\hat{H}_{F}=H-\varepsilon\hat{G}$, then
\begin{equation}
\mathcal{L}=|\braket{z|e^{iHt}e^{-i(H-\varepsilon\hat{G})t}|z}|^{2}~,
\end{equation}
By using the Zassenhaus formula $e^{t(\hat{X}+\hat{Y})}=e^{t\hat{X}}e^{t\hat{Y}}e^{-\frac{t^{2}}{2}[\hat{X},\hat{Y}]}\cdots$,
where $\hat{X}$ and $\hat{Y}$ are any operators, we can expand $e^{-it(H-\varepsilon\hat{G})}$ to rewrite 
the above expression of LE as
\begin{equation}
\mathcal{L}=|\braket{z|e^{iHt}e^{i\varepsilon\hat{G}t}e^{-iHt}e^{-\frac{\varepsilon t^{2}}{2}[\hat{G},H]}\cdots|z}|^{2}~.
\end{equation}
Finally, if we rescale the operator $\hat{G}$ in the FOTOC expression with time as $\hat{G}\rightarrow\hat{G}t$, 
we have a simple relation between the LE and the FOTOC expressions for small times,
\begin{eqnarray}
F(\hat{G}\rightarrow\hat{G}t,\,t)&=&\mathcal{L}+\frac{\varepsilon t^{2}}{2}\braket{z|
\left([\hat{G},H]+[\hat{G},H]^{\dagger}\right)|z}\notag\\
& &+\mathcal{O}(t^{3})~.\label{lefotoc}
\end{eqnarray}
By choosing $\hat{G}=\hat{J}_{x}$ Eq. (\ref{lefotoc}) takes the form
\begin{equation}
F(\hat{J}_{x}\rightarrow\hat{J}_{x}t,\,t)=\mathcal{L}+\mathcal{O}(t^{3})~,
\end{equation}
which we have also checked numerically for time $t=0.1$. In a previous analysis of the transverse XY model \cite{tapo2}, 
we had established an exponential relation between the LE and the NC. Interestingly we find a linear relation here instead
between the LE and the FOTOC operator. 

\section{NC of the FOTOC operator}
\label{ncfotoc}

The FOTOC we mentioned in earlier sections can be seen as the overlap of two states, 
and can be defined through the following process : starting from the spin coherent state $\ket{z}$, 
the first state is created by applying $\hat{V}=\ket{z}\bra{z}$, evolving for time $t$, then 
applying $\hat{W}=e^{i\varepsilon \hat{Q}}$, and then evolving for time $-t$ to get $\ket{\psi_{1}(t)}=
e^{i\hat{H}t}e^{i\varepsilon \hat{Q}}e^{-i\hat{H}t}\ket{z}$. Similarly, the second state is produced by 
evolving $\ket{z}$ for time $t$, then applying $\hat{W}$, evolving for time $-t$, and then applying 
$\hat{V}$ to get $\ket{\psi_{2}(t)}=\ket{z}\bra{z}e^{i\hat{H}t}e^{i\varepsilon \hat{Q}}e^{-i\hat{H}t}\ket{z}$. 
The FOTOC is the quantum overlap of these two states, i.e., $F(t)=\braket{\psi_{2}(t)|\psi_{1}(t)}$. 
The $e^{i\varepsilon \hat{Q}}$ operator will henceforth be referred to as the FOTOC operator and we will 
compute the NC corresponding to the time evolved FOTOC operator. We first evolve it with the Hamiltonians of 
the symmetric and the broken-symmetry phase of the excited state of the eLMG model. The Hamiltonians of the 
symmetric phase, $\hat{H}_{s}$ and broken-symmetry phase, $\hat{H}_{b}$ are \cite{GGCHV}
\begin{equation}
\hat{H}_{s}\simeq j\sqrt{1+\Omega_{x}^{2}}-\left(\frac{\sqrt{1+\Omega_{x}^{2}}-2\xi_{y}}{2}\right)
\hat{P}^{2}-\frac{\sqrt{1+\Omega_{x}^{2}}}{2}\hat{Q}^{2}~,\label{symham}
\end{equation}
and
\begin{eqnarray}
\hat{H}_{b}&\simeq& j\frac{4\xi_{y}^{2}+\Omega_{x}^{2}+1}{4\xi_{y}}-\frac{\xi_{y}\left(4\xi_{y}^{2}-\Omega_{x}^{2}
-1\right)}{4\xi_{y}^{2}-1}\hat{P}^{2}\notag\\
& &+\frac{\Omega_{x}\sqrt{4\xi_{y}^{2}-\Omega_{x}^{2}-1}}{2\left(4\xi_{y}^{2}-1\right)}
\left(\hat{Q}\hat{P}+\hat{P}\hat{Q}\right)\notag\\
& &-\frac{16\xi_{y}^{4}-8\xi_{y}^{2}+\Omega_{x}^{2}+1}{4\xi_{y}\left(4\xi_{y}^{2}-1\right)}
\hat{Q}^{2}~,\label{brokham}
\end{eqnarray}
respectively, which describe harmonic oscillators with frequencies $\omega
_{s}=(1+\Omega_{x}^{2})^{1/4}\sqrt{\Gamma_{-}}$ for the symmetric phase and $\omega
_{b}=\sqrt{4\xi_{y}^{2}-\Omega_{x}^{2}-1}$ for the broken-symmetry phase, and also 
$\Gamma_{-}=\sqrt{1+\Omega_{x}^{2}}-2\xi_{y}$. The time-dependent FOTOC operator evolved by 
$\hat{H}_{s,b}$ is $\hat{W}_{s,b}(t)=e^{i\hat{H}_{s,b}t}e^{i\varepsilon\hat{Q}}e^{-i\hat{H}_{s,b}t}$, 
and can be solved exactly using an extended version of Hadamard Lemma \cite{Zhuang} : 
$e^{\hat{A}}f(\hat{B})e^{\hat{-A}}=f(\hat{B}^{\prime})$, where $\hat{A}$ and $\hat{B}$ are any operators, 
and $f(\hat{B}^{\prime})$ is any function with
\begin{equation}
\hat{B}^{\prime}=\hat{B}+[\hat{A},\, \hat{B}]+\frac{1}{2!}[\hat{A},\,[\hat{A},\, \hat{B}]]+\cdots~.
\label{hadamard}
\end{equation}
Plugging $\hat{A}=i\hat{H}_{s,b}t$ and $\hat{B}=\hat{Q}$ in the above Eq. (\ref{hadamard}), 
we can get the exact form of the expressions of time-dependent FOTOC operators, i.e., 
$\hat{W}_{s,b}(t)=e^{i\varepsilon\hat{Q}_{s,b}(t)}$, where $\hat{Q}_{s,b}(t)=\hat{Q}
\mathcal{F}_{s,b}(t)+\hat{P}\mathcal{G}_{s,b}(t)$, with
\begin{eqnarray}
\mathcal{F}_{s}(t)&=&\cos\omega_{s}t,\quad\mathcal{G}_{s}(t)=-\frac{\sqrt{\Gamma_{-}}}
{(1+\Omega_{x}^{2})^{1/4}}\sin\omega_{s}t,\notag\\
\mathcal{F}_{b}(t)&=&\cos\omega_{b}t+\frac{\Omega_{x}\sin\omega_{b}t}{4\xi_{y}^{2}-1},
\quad\mathcal{G}_{b}(t)=-\frac{2\xi_{y}\omega_{b}\sin\omega_{b}t}{4\xi_{y}^{2}-1}~.\notag\\\label{timedepQ}
\end{eqnarray}
Now, $\hat{W}_{s,b}(t)$ are the elements, and $\{i\hat{Q},i\hat{P},-i\mathbb{1}\}$ are the generators of the 
Heisenberg group, whose algebra is defined as follows
\begin{equation}
[i\hat{Q},\,i\hat{P}]=-i\mathbb{1},\quad [i\hat{Q},\, -i\mathbb{1}]=0,\quad [i\hat{P},\, -i\mathbb{1}]=0~.
\end{equation}
The detailed calculation of computing the NC of the time-dependent FOTOC operators can be found in Appendix \ref{AppA}, 
and the final expression takes the form
\begin{equation}
\mathcal{C}_{s,b}(t)=\varepsilon^{2}\left(\mathcal{F}_{s,b}^{2}(t)+\mathcal{G}_{s,b}^{2}(t)\right)~.
\end{equation}
\begin{figure}[h!]
\centering
\includegraphics[width=0.47\textwidth]{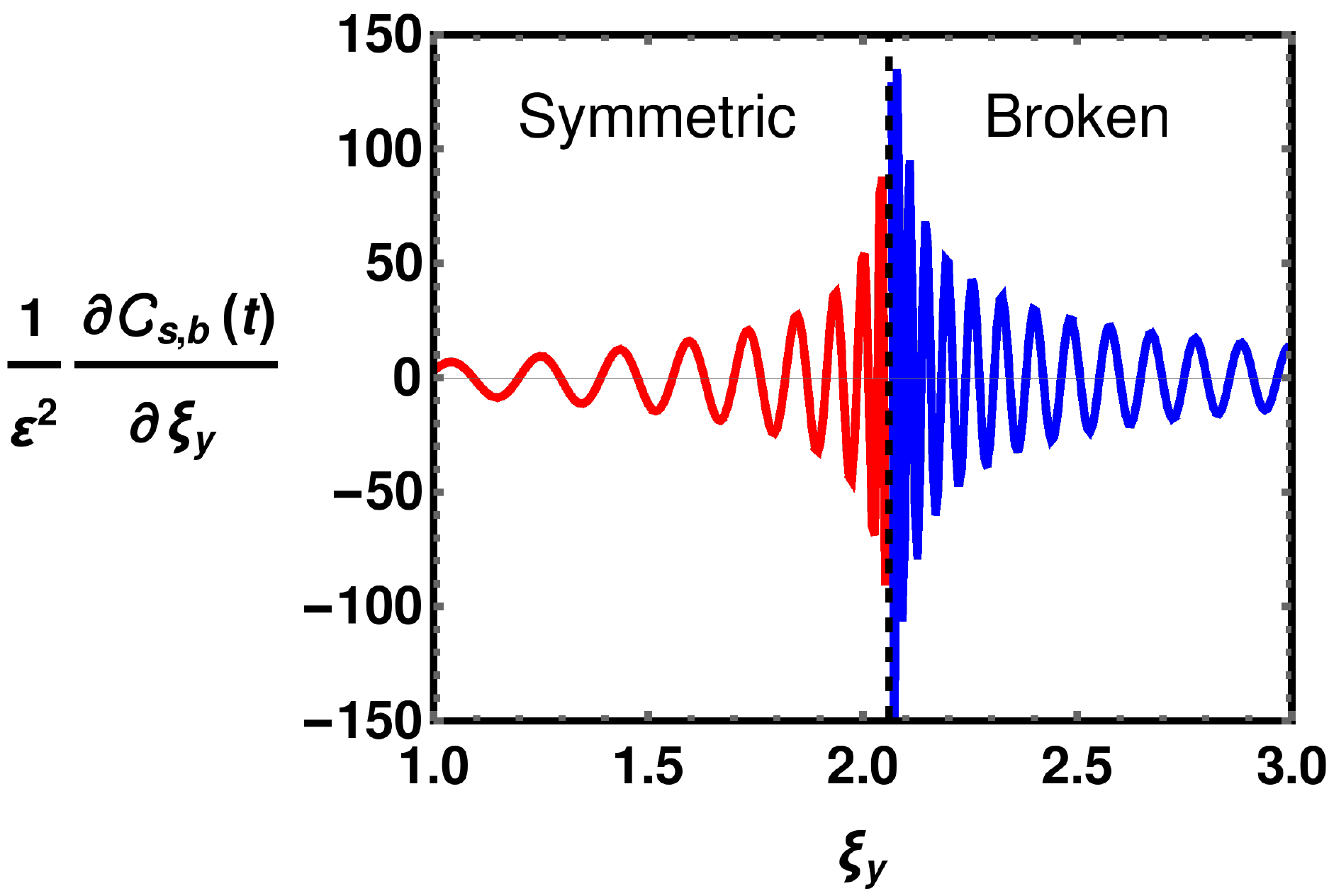}
\caption{The derivative of NC of the time-dependent FOTOC operator in the symmetric 
(solid red) and broken-symmetry (solid blue) phase for $\Omega_{x}=4$ and $t=10$.}
\label{NC}
\end{figure}
To clarify the relationship between the NC of the time-dependent FOTOC operator and the QPT, 
we first examine the dynamical behaviour of the derivative of the NC with respect to $\xi_{y}$, 
and this is shown in Fig. (\ref{NC}) for $\Omega_{x}=4$ and $t=10$. The derivative of the NC in 
the symmetric phase, $\partial\mathcal{C}_{s}(t)/\partial\xi_{y}$ shows sinusoidal behaviour, 
and the amplitude of oscillations slowly increases if we go from $\xi_{y}=0$ towards the QPT line 
$\xi_{y}=2.06$. It has a finite value at the QPT, i.e., in the limit 
$\xi_{y}\rightarrow\sqrt{1+\Omega_{x}^{2}}/2$, $\partial\mathcal{C}_{s}(t)/\partial\xi_{y}
\rightarrow2t^{2}\sqrt{1+\Omega_{x}^{2}}$, which indicates that the behaviour of the NC in the symmetric phase 
is regular. When we approach the QPT line, it remains analytical. In contrast, in the broken-symmetry phase, 
the derivative of the NC, $\partial\mathcal{C}_{b}(t)/\partial\xi_{y}$ diverges at the QPT, showing the 
non-analytical nature of the NC at these points. Furthermore, there is still sinusoidal motion in this phase, 
but the amplitude of the oscillations gradually decreases and decays to zero as we move far away from the 
QPT. In conclusion, the complexity of the FOTOC operator in the broken-symmetry phase captures information 
about QPTs, whereas this is not the case if one approaches the QPT from the symmetric phase. Although, both the phases 
resemble Hamiltonians of the harmonic oscillator, the difference comes from the presence of cross terms in $\hat{Q}$ and $\hat{P}$ 
in the broken-symmetry phase. Indeed, we may remove these cross terms by linear canonical transformations, 
but then this procedure will involve change of the basis of the system (from $\{\hat{Q},\hat{P}\}$ to $\{\hat{Q^{\prime}},\hat{P^{\prime}}\}$) 
and the comparison of the NC in the two phases will lose meaning. What we have established here is the fact that in
terms of the original $\{\hat{Q},\hat{P}\}$ coordinates, the NC serves as a diagnostic tool for the QPT in the broken-symmetry phase. 

At this point, it is important to note that the universal relation between complexity and the thermal OTOC, 
$\text{complexity}=-\log(\text{OTOC})$, mentioned in \cite{Arpan, Choudhury} does not hold with the FOTOC, as we have 
checked all the regions of the parameter space, including unstable points and QPTs. The reason for this is clear 
from the construction of the FOTOC operator, where we chose $\hat{V}$ as the projection operator rather than 
the time-independent FOTOC operator found in thermal and microcanonical OTOCs. As a result, we cannot 
establish any relationship between the FOTOC and complexity using the FOTOC operator.

\section{Geometry of quasi-scrambled excited state}
\label{qim}

The geometry of the ground and excited states of an eLMG model was studied in \cite{GGCHV}, where the 
QIM and the corresponding Ricci scalar in the parameter space $(\Omega_{x}, \xi_{y})$ was computed in 
the thermodynamic limit. The Ricci scalar turned out to be constant in the symmetric phases of both 
these states, even though a singularity appeared in all the metric components at the QPT, i.e.,
$\xi_{y}=-\sqrt{1+\Omega_{x}^{2}}/2$ for the ground state, and $\xi_{y}=\sqrt{1+\Omega_{x}^{2}}/2$ for the excited 
state although these are coordinate singularities, and not genuine curvature singularities. 
Furthermore, the Ricci scalar was found to be always negative, implying that the ground and excited state 
geometries are hyperbolic. 

Now we show that the situation qualitatively changes if we introduce a weak perturbation in the ground and excited states. 
The ground state analysis is summarised in Appendix \ref{AppB}. The wavefunction of the excited state after a
perturbation takes the form:
\begin{equation}
\ket{\Psi(t)}_{s,b}=e^{i\hat{H}_{s,b}t}e^{i\varepsilon\hat{Q}}e^{-i\hat{H}_{s,b}t}\ket{0}_{s,b}
\equiv e^{i\varepsilon\hat{Q}_{s,b}(t)}\ket{0}_{s,b}~,
\label{perturbstate}
\end{equation}
where $\hat{Q}_{s,b}$ was computed in the section \ref{ncfotoc}, 
$\hat{H}_{s,b}$ is given by Eq. (\ref{symham}) and (\ref{brokham}), and $\ket{0}_{s,b}$ is the excited 
state of the symmetric and broken-symmetry phases, with frequencies $\omega_{s}$ and $\omega_{b}$, 
respectively. Note that Eq. (\ref{QPtransform}) is only the intermediate transformation for $\hat{Q}$ 
and $\hat{P}$. The general transformation, which will cast the Hamiltonians  $\hat{H}_{s,b}$ into that of a harmonic 
oscillator is given by
\begin{eqnarray}
\hat{Q}&=&\left(\frac{\Gamma_{-}}{4\sqrt{1+\Omega_{x}^{2}}}\right)^{1/4}
\left(\gamma_{s}^{\dagger}+\gamma_{s}\right)~,\notag\\
\hat{P}&=&i\left(\frac{\sqrt{1+\Omega_{x}^{2}}}{4\Gamma_{-}}\right)^{1/4}
\left(\gamma_{s}^{\dagger}-\gamma_{s}\right)~,
\end{eqnarray}
for the symmetric phase, and $\hat{Q}=\sqrt{(\xi_{y}\omega_{b})/(4\xi_{y}^{2}-1)}(\gamma_{b}^{\dagger}+\gamma_{b})$,
$\hat{P}=\mathcal{V}\gamma_{b}^{\dagger}+\mathcal{V}^{\ast}\gamma_{b}$, for the broken-symmetry phase, where 
\begin{equation}
\mathcal{V}=\frac{\Omega_{x}}{\sqrt{4\xi_{y}\omega_{b}(4\xi_{y}^{2}-1)}}+
i\sqrt{\frac{4\xi_{y}^{2}-1}{4\xi_{y}\omega_{b}}}~.
\end{equation}
\begin{figure}[h!]
\centering
\includegraphics[width=0.47\textwidth]{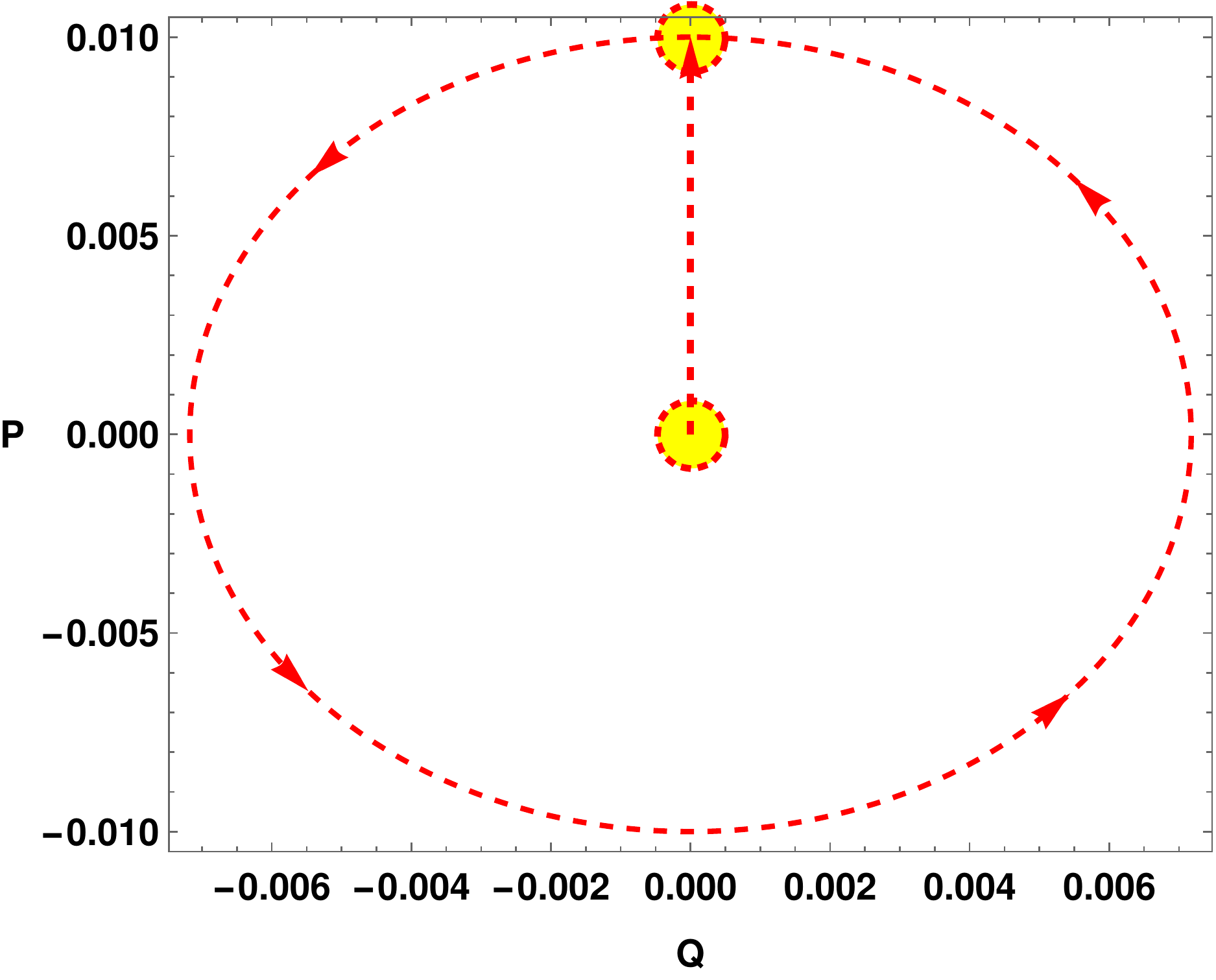}
\caption{Phase space distribution of the FOTOC operator with $\Omega_{x}=4$, $\xi_{y}=1$, and $\varepsilon=0.01$. 
The dotted circle represents the quantum uncertainty of the ground state, which moves around in phase space with time, 
as shown by dotted lines and arrows. The distribution (uncertainty bubble) remains localized under Gaussian 
dynamics $\hat{H}_{s,b}$, depicting quasi-scrambling.}
\label{phasetravel}
\end{figure}
Here, $\mathcal{V}^{\ast}$ denotes the complex conjugate of $\mathcal{V}$, and $\gamma_{s, b}^{\dagger}$ and 
$\gamma_{s, b}$ are the Bogoliubov creation and annihilation operators respectively, such that 
$\gamma_{s, b}\ket{0}_{s,b}=0$. In \cite{Shenker, Susskind}, a nice physical interpretation 
of the perturbed state of the form Eq. (\ref{perturbstate}) was given : we first prepare the excited 
state at time $t=0$, and then, at a later time $t$, we apply a suitably small perturbation, i.e., a 
FOTOC operator, to obtain the perturbed excited state. Eq. (\ref{perturbstate}) can now be read 
straightforwardly as follows : for a given excited state, the operator $e^{-i\hat{H}_{s,b}t}$ acts trivially 
on that state. Then, the FOTOC operator is a kind of displacement operator that shifts the state by 
momentum $\varepsilon$ in phase space. Finally, we perform the reverse time evolution, which typically results 
in a state's rotation about the origin of the phase space. Fig. (\ref{phasetravel}) depicts a visual representation 
of these phase space distributions, with the ground state characterised by an uncertainty bubble. The graphic 
clearly shows that the distribution remains confined, but has traveled around in phase space, a phenomenon 
referred to in the current literature as quasi-scrambling \cite{Zhuang}.

As mentioned in the introduction, there are two distinct types of scramblings : genuine scrambling, in which the distribution 
of operators spreads widely throughout the phase space, and quasi-scrambling, in which the distribution stays 
localised but is mobile in phase space. One of the key distinctions between quasi-scrambling and genuine 
scrambling is whether the time evolution is generated by Gaussian or non-Gaussian unitaries. Gaussian dynamics 
does not have the ability to spread an operator distribution in phase space, and we identify its dynamics 
as quasi-scrambling. On the other hand, in non-Gaussian dynamics, an operator distribution spreads significantly 
over phase space, and thus non-Gaussian dynamics give genuine scrambling. In Eq. (\ref{perturbstate}), the 
temporal evolution is performed by Gaussian unitaries generated by the Hamiltonians $\hat{H}_{s, b}$. So, the 
perturbed state $\ket{\Psi(t)}_{s,b}$ is a state after the information has been quasi-scrambled, and we will 
now study the geometry of this quasi-scrambled state. 

Following \cite{tapo1, tapo2, tapo3, tapo4}, 
the QIM $g_{ij}$ is the real (symmetric) part of quantum geometric tensor denoted by $\chi_{ij}$ takes the form 
\begin{equation}
\chi_{ij}=\sum\limits_{m\neq n}\frac{\braket{n|\partial_{i}\hat{H}|m}
\braket{m|\partial_{j}\hat{H}|n}}{\left(E_{m}-E_{n}\right)^{2}}~,
\end{equation}
where $E_{n}$ is the non-degenerate eigenvalue corresponding to the orthonormal eigenvector $\ket{n}$ of 
the system whose Hamiltonian is $\hat{H}$. To study the geometry of the perturbed state, we first rewrite 
it in a simpler form using Bogoliubov operators $\gamma_{s,b}$, and further using the Baker-Campbell-Hausdorff 
formula, we obtain
\begin{equation}
\ket{\Psi(t)}_{s,b}=\sum\limits_{l=0}^{\infty}e^{-\frac{\varepsilon^{2}}{2}|\mathcal{B}_{s,b}(t)|^{2}}
\frac{\left(i\varepsilon\mathcal{B}_{s,b}(t)\right)^{l}}{\sqrt{l!}}\ket{l}_{s,b}~,
\end{equation}
where we have defined
\begin{eqnarray}
\mathcal{B}_{s}(t)&=&\left(\frac{\Gamma_{-}}{4\sqrt{1+\Omega_{x}^{2}}}\right)^{1/4}
\mathcal{F}_{s}(t)+i\left(\frac{\sqrt{1+\Omega_{x}^{2}}}{4\Gamma_{-}}\right)^{1/4}\mathcal{G}_{s}(t)~,\notag\\
\mathcal{B}_{b}(t)&=&\sqrt{\frac{\xi_{y}\omega_{b}}{4\xi_{y}^{2}-1}}\mathcal{F}_{b}(t)+\mathcal{V}\mathcal{G}_{b}(t)~,
\end{eqnarray}
and $\ket{l}_{s,b}$ are the eigenstates of $\gamma_{s,b}$. We can expand
\begin{equation}
\ket{\Psi(t)}_{s,b}=\ket{0}_{s,b}+i\varepsilon\mathcal{B}_{s,b}(t)\ket{1}_{s,b}+\mathcal{O}(\varepsilon^{2})~,
\end{equation} 
since $\varepsilon$ is a small number, which we will choose to be $\varepsilon=0.01$ in the analysis to follow. 
The QIM for the state $\ket{0}_{s,b}$ has been computed in \cite{GGCHV}, and we list the additional
terms, proportional to $\varepsilon$ in the symmetric phase, 
\begin{eqnarray}
g_{\Omega_{x}\Omega_{x}}&=&\varepsilon\frac{\sqrt{j}\Omega_{x}t\left(2\xi_{y}^{2}-3\xi_{y}\sqrt{1+\Omega_{x}^{2}}+
1+\Omega_{x}^{2}\right)\cos\omega_{s}t}{(1+\Omega_{x}^{2})^{7/4}\Gamma_{-}^{3/2}}~,\notag\\
g_{\Omega_{x}\xi_{y}}&=&-\varepsilon\frac{\sqrt{j}\,t\cos\omega_{s}t}{2(1+\Omega_{x}^{2})^{3/4}\Gamma_{-}^{1/2}},
\quad g_{\Omega_{x}t}=\varepsilon\frac{\sqrt{j}\,\Gamma_{-}^{1/2}\cos\omega_{s}t}{2(1+\Omega_{x}^{2})^{3/4}}~,\notag\\
g_{\xi_{y}\xi_{y}}&=&g_{\xi_{y}t}=g_{tt}=0~, \label{metric}
\end{eqnarray}
\begin{figure}[h!]
\centering
\includegraphics[width=0.47\textwidth]{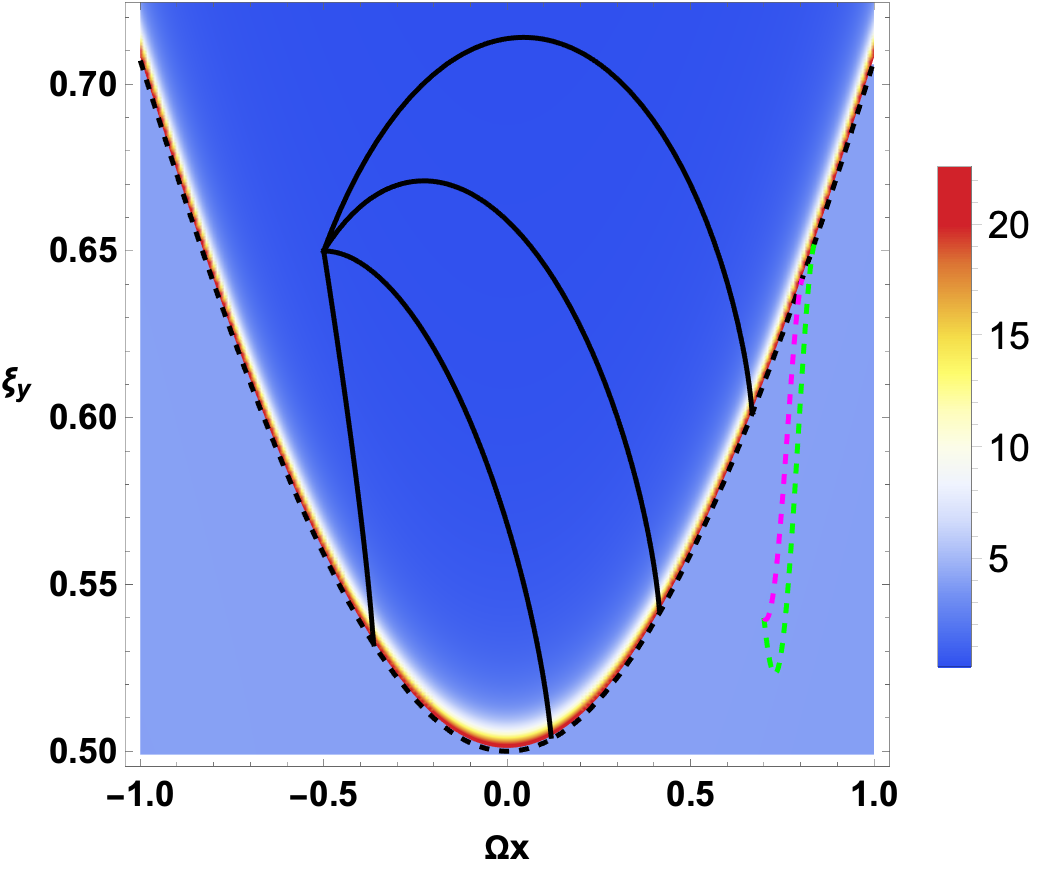}
\caption{The Ricci scalar of the geometry of the quasi-scrambled state, for $t=4$, $j=100$. The dashed
black line separates the two phases. The solid black, the green and the magenta lines are plots of
geodesics, see the main text.}
\label{FigLMG}
\end{figure}
where $\mathcal{O}(1/j)$ terms are neglected. It becomes cumbersome to present the equations 
of the metric components in the broken-symmetry phase, and we omit these for brevity. 

Our main results are shown graphically in Fig. (\ref{FigLMG}), where the background depicts the 
Ricci scalar in the two phases separated by the dashed black line. In the broken-symmetry phase
$\xi_y>1/2\sqrt{1+\Omega_x^2}$, this diverges at the phase transition, and we find that this divergence
is absent if we approach the phase transition from the symmetric phase. Here, for illustration, we have taken $t=4$ and 
$j=100$, with $\varepsilon=0.01$ as mentioned before. We also plot a few typical geodesics for the geometries, 
by numerically solving the geodesic equation 
\begin{equation}
\frac{d^{2}x^{i}}{d\tau^{2}}+\Gamma^{i}_{jl} \frac{dx^{j}}{d\tau} \frac{dx^{l}}{d\tau}=0~,
\end{equation}
with $\Gamma^{i}_{jl}$ being the Christoffel connections, and $x^{i}=(\Omega_{x}(\tau),\xi_{y}(\tau))$,
and we use the normalisation condition $g_{ij}\dot{x}^{i}\dot{x}^{j}=1$. Here $\tau$ is an affine parameter along the geodesic
and the overdot indicates a differentiation with respect to $\tau$. In the broken-symmetry phase, we
start the geodesics from $(\Omega_x,\xi_y)=(-0.5,0.65)$ with the initial value of ${\dot \xi_y} =-0.08, 0, 0.02$ and $0.03$ for the 
solid black lines from bottom to top, respectively, and the initial value of ${\dot \Omega_x}$ is determined from the normalisation condition. 
In the symmetric phase, the dashed green and magenta lines correspond to the initial value of 
$(\Omega_x,\xi_y)=(0.7,0.54)$ with the initial value of ${\dot \xi_y} = -0.12$ and $-0.02$, respectively. 

In the symmetric phase, our results indicate that $R\to -4$ as $\xi_y \to 1/2\sqrt{1+\Omega_x^2}$. Away from 
this phase transition line, the dependence of $R$ with $\Omega_x$ and $\xi_y$ does not show any interesting behaviour. 
We also computed the Fubini-Study complexity (defined as the geodesic length) by numerically inverting the solution of 
the geodesic equation \cite{tapo1, tapo2, tapo3, tapo4}. It can be seen from Fig. (\ref{FigLMG}) that 
In both phases, the geodesics are ``attracted'' towards the phase boundary and ends there, and these 
do not show any particularly interesting behaviour.   

\section{Conclusions}
In this work, we have demonstrated how FOTOCs have distinctive behaviour at QPTs in the ground and excited states. 
We have illustrated this by analysing the dynamics of the FOTOC in two phases of an eLMG model. We have also established 
the connection between the FOTOC and the Loschmidt echo at small times, by a suitable time-rescaling. 
Next, we computed the NC of the FOTOC operator and showed that its derivative is divergent 
as one approaches QPT from the broken-symmetry phase side. Also, the geometry of the parameter space of this model was 
studied by perturbing the ground and excited states with the FOTOC operator. These states are quasi-scrambled 
states after the perturbation, and analytic computations were performed to extract the QIM. This yields a diverging Ricci scalar at QPTs 
from the broken-symmetry phase side, in contrast to the unperturbed states whose Ricci scalar does not possess any such feature.

Our study focussed only on the eLMG model, whose Hamiltonian reduces to that of a harmonic oscillator : a Gaussian unitary.
The time-evolution generated by a Gaussian unitary will give quasi-scrambling. It will be interesting to explore the QIG with 
non-Gaussian unitaries that offer genuine scrambling. We leave this for a future work. 

\appendix
\section{}
\label{AppA}
In this section, we present a detailed calculation of the NC following \cite{Arpan}. We apply 
Nielsen's geometric approach to find the optimal circuit. The first step is to parametrize the unitary 
as a path-ordered exponential
\begin{equation}
U(\tau)=\overleftarrow{\mathcal{P}}\exp\left(\int_{0}^{\tau}d\tau^{\prime}H(\tau^{\prime})\right)~,
\label{path}
\end{equation}
with $H(\tau^{\prime})=\sum_{I\in \{1,3\}}Y^{I}(\tau^{\prime})M_{I}$, $Y^{I}(\tau^{\prime})$ are control 
functions and specifies a particular circuit in the space of unitaries, and $M_{I}$ are the 
three-dimensional representation of the Heisenberg group generators
\begin{equation}
M_{1}=
\begin{bmatrix}
0&1&0\\
0&0&0\\
0&0&0\\
\end{bmatrix}
,\quad M_{2}=
\begin{bmatrix}
0&0&0\\
0&0&1\\
0&0&0\\
\end{bmatrix}
,\quad M_{3}=
\begin{bmatrix}
0&0&1\\
0&0&0\\
0&0&0\\
\end{bmatrix}~,
\end{equation}
satisfying the Heisenberg algebra. The explicit form of the functions $Y^{I}(\tau)$ can 
be evaluated by using Eq. (\ref{path}) and the expressions of $M_{I}$,
\begin{equation}
Y^{I}(\tau)=-\text{Tr}\left[(\partial_{\tau}U(\tau)) U^{-1}(\tau)M_{I}^{T}\right].~
\end{equation}
By following the usual procedure \cite{Liu}, we can define the cost functional for various paths 
\begin{equation}
\mathcal{D}[U(\tau)]=\int_{0}^{1}d\tau^{\prime}\sum_{I}|Y^{I}(\tau^{\prime})|^{2},~\label{cost}
\end{equation}
where we followed $\kappa=2$ cost functions. Note that the $\kappa=2$ cost function and the $F_{2}$ cost 
function will provide the exact same extremal trajectories or optimal circuits \cite{Guo}. The minimal value 
of the above cost functional will give the required complexity, and it can be obtained by evaluating it on 
the geodesics of unitary space with the boundary conditions
\begin{equation}
\tau=0,\,U(\tau=0)=\mathbb{1},\quad \tau=1,\,U(\tau=1)=\hat{W}_{s,b}(t).~
\end{equation}
We first represent the FOTOC operator in Heisenberg group generators as $e^{\varepsilon
\left(\mathcal{F}_{s,b}(t)M_{1}+\mathcal{G}_{s,b}(t)M_{2}\right)}$, which is solved as,
\begin{equation}
\hat{W}_{s,b}(t)=
\begin{bmatrix}
1&\varepsilon\mathcal{F}_{s,b}(t)&\frac{\varepsilon^{2}}{2}\mathcal{F}_{s,b}(t)\mathcal{G}_{s,b}(t)\\
0&1&\varepsilon\mathcal{G}_{s,b}(t)\\
0&0&1\\
\end{bmatrix}~.
\end{equation}
In general, we can parametrize an element of Heisenberg group by $U(\tau)$ where,
\begin{equation}
U(\tau)=\begin{bmatrix}
1&x_{1}(\tau)&x_{3}(\tau)\\
0&1&x_{2}(\tau)\\
0&0&1\\
\end{bmatrix}~,
\end{equation}
then the Eq. (\ref{cost}) can be written in the form:
\begin{eqnarray}
\mathcal{D}[U(\tau)]&=&\int_{0}^{1}d\tau\Bigg((1+x_{2}^{2})\left(\frac{dx_{1}}{d\tau}\right)^{2}+
\left(\frac{dx_{2}}{d\tau}\right)^{2}+\notag\\
& &\left(\frac{dx_{3}}{d\tau}\right)^{2}-2x_{2}\left(\frac{dx_{1}}{d\tau}\right)
\left(\frac{dx_{3}}{d\tau}\right)\Bigg)~.\label{unitarymetric}
\end{eqnarray}
The geodesic equations corresponding to the above metric are 
\begin{eqnarray}
\dot{x}_{2}(x_{2}\dot{x}_{1}-\dot{x}_{3})+\ddot{x}_{1}=0,& &\dot{x}_{1}(-x_{2}\dot{x}_{1}-
\dot{x}_{3})+\ddot{x}_{2}=0,\notag\\
\dot{x}_{2}((x_{2}^{2}-1)\dot{x}_{1}-x_{2}\dot{x}_{3})+\ddot{x}_{3}&=&0~,
\end{eqnarray}
where the dot represents the derivative with respect to $\tau$, and is solved using above boundary conditions to get
\begin{eqnarray}
x_{1}(\tau)&=&\varepsilon\mathcal{F}_{s,b}(t)\tau,\quad x_{2}(\tau)=\varepsilon\mathcal{G}_{s,b}(t)\tau,\notag\\
x_{3}(\tau)&=&\frac{\varepsilon^{2}}{2}\mathcal{F}_{s,b}(t)\mathcal{G}_{s,b}(t)\tau^{2}.~
\end{eqnarray}
Using the above solution in Eq. (\ref{unitarymetric}), the final expression for the NC is 
\begin{equation}
\mathcal{C}_{s,b}(t)=\varepsilon^{2}\left(\mathcal{F}_{s,b}^{2}(t)+\mathcal{G}_{s,b}^{2}(t)\right)~.
\end{equation}

\section{}
\label{AppB}
In this section, for the sake of completeness, we evaluate the metric components of the perturbed ground state. 
The Hamiltonian of the ground state takes the form:
\begin{equation}
\hat{H}_{g}\simeq -j\sqrt{1+\Omega_{x}^{2}}+\left(\frac{\sqrt{1+\Omega_{x}^{2}}+2\xi_{y}}{2}\right)\hat{P}^{2}+
\frac{\sqrt{1+\Omega_{x}^{2}}}{2}\hat{Q}^{2}~,
\end{equation}
which describes the harmonic oscillator with frequency $\omega_{g}=(1+\Omega_{x}^{2})^{1/4}\sqrt{\Gamma_{+}}$ 
and $\Gamma_{+}=\sqrt{1+\Omega_{x}^{2}}+2\xi_{y}$. The Bogoliubov transformation of the ground state is
\begin{eqnarray}
\hat{Q}&=&\left(\frac{\Gamma_{+}}{4\sqrt{1+\Omega_{x}^{2}}}\right)^{1/4}
\left(\gamma_{g}^{\dagger}+\gamma_{g}\right)~,\notag\\
\hat{P}&=&i\left(\frac{\sqrt{1+\Omega_{x}^{2}}}{4\Gamma_{+}}\right)^{1/4}
\left(\gamma_{g}^{\dagger}-\gamma_{g}\right)~,
\end{eqnarray}
with $\gamma_{g}\ket{0}_{g}=0$ and $\ket{0}_{g}$ is the Bogoliubov ground state. Following a similar 
analysis as for the excited state, we perturb the ground state as
\begin{equation}
\ket{\Psi(t)}_{g}=e^{i\hat{H}_{g}t}e^{i\varepsilon\hat{Q}}e^{-i\hat{H}_{g}t}\ket{0}_{g}\equiv 
e^{i\varepsilon\hat{Q}_{g}(t)}\ket{0}_{g}~,
\end{equation}
with 
\begin{equation}
\hat{Q}_{g}=\hat{Q}\cos\omega_{g}t+\hat{P}\,\frac{\Gamma_{+}^{1/2}\sin\omega_{g}t}{(1+\Omega_{x}^{2})^{1/4}}~,
\end{equation}
and the metric components proportional to $\varepsilon$ turn out to be
\begin{eqnarray}
g_{\Omega_{x}\Omega_{x}}&=&\varepsilon\frac{\sqrt{j}\,\Omega_{x}t\left(2\xi_{y}^{2}+3\xi_{y}
\sqrt{1+\Omega_{x}^{2}}+1+\Omega_{x}^{2}\right)\cos\omega_{g}t}{(1+\Omega_{x}^{2})^{7/4}\Gamma_{+}^{3/2}}\notag\\
& &-\varepsilon\frac{\sqrt{j}\,\Omega_{x}\xi_{y}\sin\omega_{g}t}{(1+\Omega_{x}^{2})^{2}\Gamma_{+}}~,\notag\\
g_{\Omega_{x}\xi_{y}}&=&\varepsilon\frac{\sqrt{j}\,t\cos\omega_{g}t}{2(1+\Omega_{x}^{2})^{3/4}\Gamma_{+}^{1/2}}+
\varepsilon\frac{\sqrt{j}\,\sin\omega_{g}t}{2(1+\Omega_{x}^{2})\Gamma_{+}}~,\notag\\
\quad g_{\Omega_{x}t}&=&\varepsilon\frac{\sqrt{j}\,\Gamma_{+}^{1/2}\cos\omega_{g}t}
{2(1+\Omega_{x}^{2})^{3/4}},\quad g_{\xi_{y}\xi_{y}}=g_{\xi_{y}t}=g_{tt}=0~.
\end{eqnarray}
We have computed the Ricci scalar in the parameter space $(\Omega_{x},\xi_{y})$ for fixed $t$ and find that $R\to -4$ 
as $\xi_y \to -1/2\sqrt{1+\Omega_x^2}$ similar to the situation in the symmetric phase.

\end{document}